\documentclass[aps,prb,twocolumn,showpacs,superscriptaddress]{revtex4-1}
\usepackage{color}
\usepackage{graphicx}
\usepackage{dcolumn}
\usepackage{amsmath}
\usepackage{amssymb}
\usepackage{subfigure,amsmath,verbatim,moreverb,bm}
\def\be{\begin{equation}}
\def\ee{\end{equation}}
\def\ber{\begin{eqnarray}}
\def\eer{\end{eqnarray}}

\def\xv{{\bf x}}
\def\rv{{\bf r}}

\def\pv{{\bf p}}

\def\dv{{\bf d}}

\newcommand{\unit}{\mathbb{\hat I}}
\def\id{{\rm i}}

\def\nn{\nonumber}

\def \SE {Schr\"odinger equation~}

\def	 \Q {Eq.~}
\def \d {\mathrm{d}}

\def\Im{{\rm Im}}
\def\Re{{\rm Re}}

\begin{document}
\title{Quantum electrodynamical time-dependent density functional theory on a lattice}
\author{M. Farzanehpour}
\email{m.farzanehpour@gmail.com}
\affiliation{Nano-Bio Spectroscopy group and ETSF Scientific Development Centre, 
  Departamento de F\'isica de Materiales, Universidad del Pa\'is Vasco UPV/EHU, E-20018 San Sebasti\'an, Spain}
\author{ I. V. Tokatly}
\email{ilya.tokatly@ehu.es}
\affiliation{Nano-Bio Spectroscopy group and ETSF Scientific Development Centre, 
  Departamento de F\'isica de Materiales, Universidad del Pa\'is Vasco UPV/EHU, E-20018 San Sebasti\'an, Spain}
\affiliation{IKERBASQUE, Basque Foundation for Science, E-48011 Bilbao, Spain}
\date{\today}

\begin{abstract}
We present a rigorous formulation of the time-dependent density functional theory for interacting lattice electrons strongly coupled to cavity photons. We start with an example of one particle on a Hubbard dimer coupled to a single photonic mode, which is equivalent to the single mode spin-boson model or the quantum Rabi model. For this system we prove that the electron-photon wave function is a unique functional of the electronic density and the expectation value of the photonic coordinate, provided the initial state and the density satisfy a set of well defined conditions. Then we generalize the formalism to many interacting electrons on a lattice coupled to multiple photonic modes and prove the general mapping theorem. We also show that for a system evolving from the ground state of a lattice Hamiltonian any density with a continuous second time derivative is locally $v$-representable. 
\end{abstract}
\pacs{31.15.ee, 71.10.Fd} 
\maketitle

\section{Introduction}
Time-dependent density functional theory (TDDFT) is a formulation of quantum many-body problem based on the one-to-one map from the time-dependent density to the driving external potential \cite{RunGro1984,Ulrich-book,TDDFTBook2012}. The main practical outcome of this map, which underlies the popularity of TDDFT, is a possibility to reproduce the dynamics of the exact density in the interacting many-body system by solving an auxiliary problem for non-interacting Kohn-Sham (KS) system driven by a properly adjusted self-consistent potential.  This dramatically reduces  the complexity of the problem and necessary computational resource  to study the dynamics of many-particle systems.

The standard TDDFT assumes that the system is driven by a classical time-dependent electromagnetic field \cite{Ulrich-book,TDDFTBook2012}. This approach is indeed sufficient for most typical situations in quantum chemistry and condensed matter physics. However, in the recent years, with the impressive progress in the fields of cavity and circuit quantum electrodynamics (QED) it has been made possible to experimentally study systems interacting  strongly with quantum light, like atoms in optical cavity \cite{MabuchiDoherty2002,RaiBruHar2001,waltheretalIOP2006},
superconducting qubits and quantum dots \cite{Blaisetall2004,WallraffetalNat2004,SorCleGel2004,YouNoriNature2011} , trapped ions \cite{Leibriedetal2003} and  molecules interacting with cavity photons \cite{Schowartzetal2011,Hutchisonetal2012,MorralStellacciNature2012}.

Recently a generalization of TDDFT for quantum many-electron systems coupled to cavity photons has been proposed \cite{TokatlyPrl2013,Ruggenthaler2etal2014}. This theory, which can be named QED-TDDFT, relies on a generalized mapping theorem stating that there exists a unique map from the time-dependent electronic  density and the expectation value of the photonic filed to the total electron-photon wave function. In Ref.~\onlinecite{TokatlyPrl2013} the uniqueness of this generalized mapping has been demonstrated using the Taylor expansion technique  under the standard in TDDFT assumption of analyticity in time \cite{vanLeeuwen1999,RunGro1984}.  The question of existence of the density-to-potential map, which is known in DFT as the $v$-representability problem, is much more difficult. In fact, in the standard TDDFT for continuum systems this question is still not fully resolved, although a significant progress has been made recently \cite{RugLee2011}.  

Currently a complete and rigorous formulation of TDDFT, including the resolution of the $v$-representability problem, is available for lattice many-body systems \cite{TokatlyL2011,FarTok2012}. The aim of the present paper is to extend the uniqueness and existence theorems of the lattice TDDFT to systems strongly interacting with a quantized electromagnetic field. 

To make the idea of the proof more transparent we start with the simplest nontrivial system of one electron on a two-site lattice (a Hubbard dimer) coupled to a single photonic mode. It is worth noting that formulation of TDDFT for this system has it own value. Indeed, the dimer coupled to a quantum Bose field is (unitary) equivalent to such well known and popular models as the quantum Rabi model and the spin-boson model \cite{Rabi1936,Leggettetal1987,BraakPRL2011} which have a wide variety of applications ranging from  quantum optics \citep{Vedral2005} and molecular physics  \cite{ThaPasKis2004} to the magnetic resonance  in solid state physics \cite{IrishPrl2007}. There is also a natural connection to the Dicke model \cite{Dicke1954,chenetal2008}.  For this system we prove that, provided some well-defined conditions are fulfilled, there exists a unique mapping from the time-dependent on-site density  and the expectation value of the bosonic coordinate to the wave function and the external deriving  
potentials.  Afterwards  we extend the QED-TDDFT mapping theorem to the general case of $N$ interacting electrons on an $M$-site lattice coupled to multiple photonic modes. We also prove that, similarly to the standard lattice TDDFT \cite{FarTok2012}, the local existence/$v$-representablity is guaranteed if the dynamics starts from the ground state of a lattice Hamiltonian.

The structure of the paper is the following. In Sec.~\ref{two-level} we present a complete formulation of QED-TDDFT for the Hubbard dimer coupled to a single photonic mode. We derive the equation of motion for the expectation value of the field and the force balance equation and construct the corresponding universal nonlinear \SE (NLSE). Then we prove the maping theorem of QED-TDDFT for this model by applying the known results from the theory of semilinear partial differential equations (PDE) \cite{Segal1963,Haraux}. In Sec.~\ref{many-electron-photon} we generalize the formalism to the system of many particles on a many-site lattice which is coupled to multiple photonic modes. We derive the corresponding NLSE the many-body system and then formulate and prove the general existence and uniqueness theorem for the lattice QED-TDDFT.  Section~\ref{ground-state-theorem}  presents a practically important case of a system evolving from its ground state. The main outcome of this section is  the theorem of a 
local $v$-representability for the initial ground state. In Conclusion we summarize our results.

%###################################################################################################
\section{QED-TDDFT for a Hubbard dimer coupled to a single photonic mode}\label{two-level}
%###################################################################################################

To make our approach more transparent and clear we  consider first a simple system one quantum particle on a two-site lattice, which is coupled to a single-mode photonic field. The state of the system at time $t$ is characterized by the electron-photon wave function $\psi_i(p;t)=\langle i,p|\Psi(t)\rangle$, where the index $i=\{1,2\}$ corresponds to the particle ``coordinate'' and takes values on the lattice sites, and the real continuum variable $p$ describes the photonic degree of freedom. 

\begin{figure}
\begin{centering} 
\includegraphics[height=0.23\textwidth]{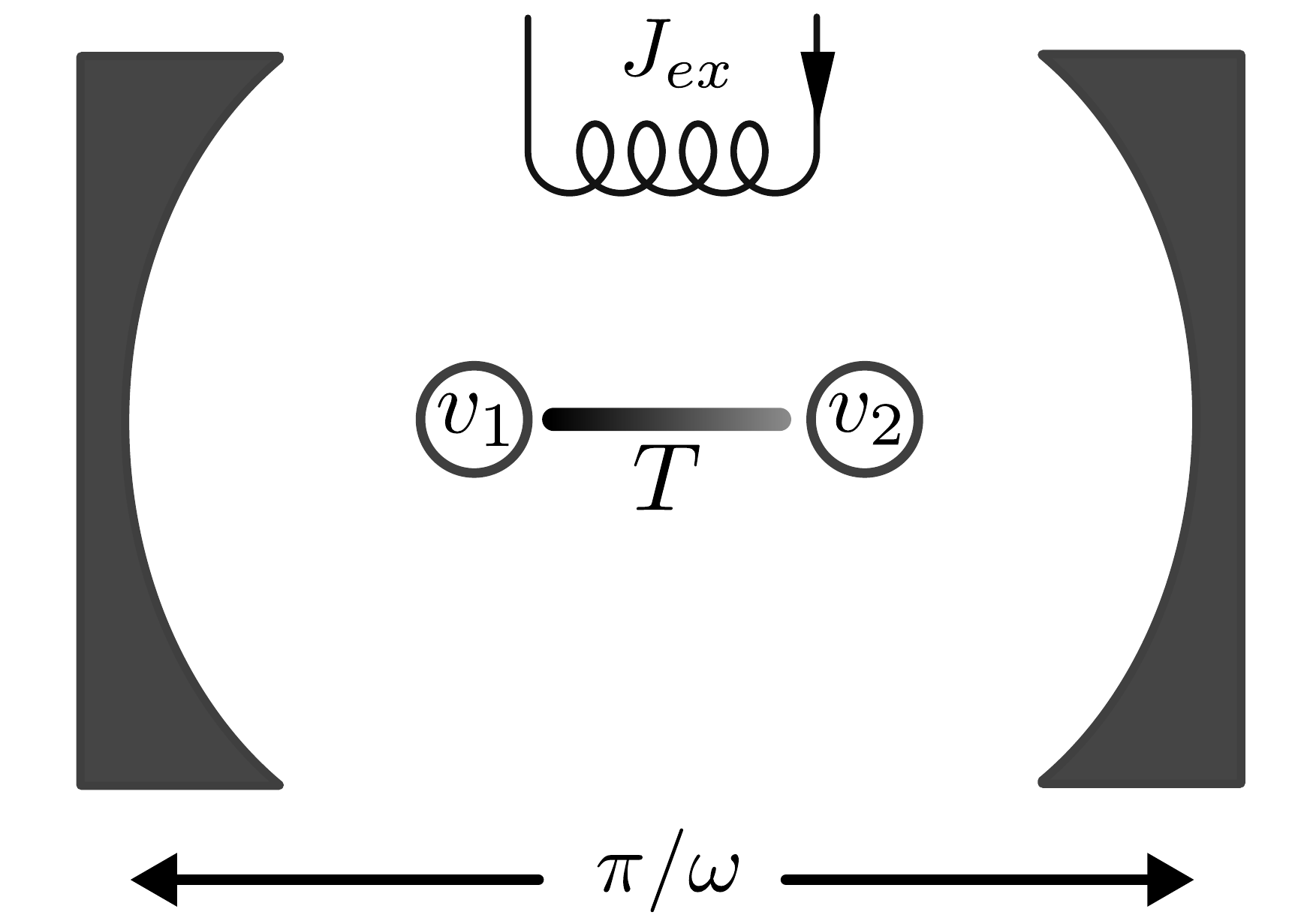}
\caption{ A schematic view of a two-site lattice in a cavity. The electron, which can tunnel from one site to the other with the hopping rate $T$, experiences the on-site potential $v_i(t)$  specific to that site. The photonic field in the cavity is driven by a time-dependent external current $J_{ex}(t)$. The wave length of the electromagnetic field $2\pi/\omega$ is proportional to the cavity size and assumed to be much larger than the lattice size, so that we can adopt the dipole approximation for the electron-photon interaction. }
\label{fig:cavity}
\end{centering}
\end{figure}

In this model the electronic density is coupled to the external on-site potential $v_i(t)$  which acts on the individual sites, while the photonic subsystem can be driven (exited) independently by an external time-dependent ``current''  $J_{ex}(t)$.  Assuming that the wavelength of the photon field is much larger than the size of the system, we adopt the dipole approximation. Figure \ref{fig:cavity} shows a schematic view of a two-site lattice in a quantum cavity.

The following time-dependent \SE governs the time evolution of the electron-photon wave function $\psi_i(p;t)$ from a given  initial state $\psi_i(p,t_0)$
\begin{subequations} \label{SE1}
\begin{eqnarray}
\id&&  \partial_t \psi_1(p;t)= - T \psi_2(p;t)\\
 &&+\left(\frac{- \partial_p ^2}{2} + \frac{\omega^2 p^2}{2}+J_{ex}(t)p +\lambda p +v_1 (t)\right) \psi_1(p;t),\nn\\
\id&& \partial_t \psi_2(p;t)=  - T \psi_1(p;t)\\
 &&+ \left(\frac{- \partial_p ^2}{2} + \frac{\omega^2 p^2}{2}+J_{ex}(t)p -\lambda p +v_2(t)\right)\psi_2(p;t),\nn
\end{eqnarray}
\end{subequations}
where the real coefficient $T$ corresponds to the rate of hopping from one site to the other, $\omega$ is the frequency of the photon mode and $\lambda$ is the electron-photon coupling constant (see figure \ref{fig:cavity}). 
%%%%%%%%%%%%%%%%%%%%%%%%%
%                           Figure                                %
%%%%%%%%%%%%%%%%%%%%%%%%%

Formally Eqs.~(\ref{SE1}) describes a driven two-level system coupled to a quantum harmonic oscillator.  Because of the gauge invariance the physics is not changed if we add to the potential a global time-dependent constant. Therefore without loss of generality we can adopt the gauge condition $v_1+v_2=0$ and define the on-site potential as follows $v(t)=v_1(t)=-v_2(t)$. With this definition the Hamiltonian in Eq.~(\ref{SE1}) takes to form
% \begin{equation}\label{SE2}
% \id \partial_t \Psi(t)= \hat H(t) \Psi(t),
% \end{equation}
\begin{equation}\label{H}
%\begin{split}
\hat H(t)= - T  \hat \sigma_x+v(t) \hat \sigma_z+\lambda p \hat \sigma_z
-\frac{\partial _p^2}{2}  +\frac{ \omega ^2 p^2}{2} +J_{ex}(t)p.
%\end{split}
\end{equation}
$\hat \sigma_x$, and $\hat \sigma_z$ are the Pauli matrices, and a $2\times 2$ unit matrix is assumed in the last three terms. The first two terms in Eq.~(\ref{H}) correspond to a two-level system (spin 1/2), while the last three terms describe a driven harmonic oscillator. Finally the third term in Eq.~(\ref{H}) is a linear coupling between the discrete and continuous variables.  It is now clear the Eq.~\eqref{SE1} is equivalent to the \SE for the Rabi model or single mode spin-boson model\cite{Rabi1936,Leggettetal1987,BraakPRL2011}. Therefore the subsequent discussion and all results of this section are directly applicable to these models. We also note that a detailed derivation of Hamiltonian Eq.~(\ref{H}) for a nonrelativistic system in a cavity can be found in Ref.~\onlinecite{Ruggenthaler2etal2014} (see Appendix~E).

Let us now turn to the formulation of QED-TDDFT. In general all DFT-like approaches assume that the state of the system is  uniquely determined by small set of basic observables, such as the density in TDDFT, the current in TDCDFT.  Below we prove a theorem which generalizes the lattice-TDDFT of Ref.~\onlinecite{FarTok2012} to the system coupled to a quantum oscillatoric degree of freedom as defined in Eq.~(\ref{SE1}). Namely, we will prove that, provided some well defined conditions are fulfilled, the electron-photon wave function $\Psi(t)$ is uniquely determined by the on-site density $n_{i}$ and the expectation value of the photonic coordinate $P=\langle p \rangle$.

In our formulation we follow the NLSE approach to TDDFT\cite{Tokatly2007,Maitra2010,TokatlyUni2011,TokatlyL2011,FarTok2012} and adopt the same general logic as in Ref.~\onlinecite{FarTok2012}. We start with defining the basic observables for our two-site model. The first basic variable,  is the on-site density $n_i$ -- the number of particles on  the site $i$ 
\begin{equation} \label{D1}
 n_i(t)= \int \d p |\psi_i(p;t)|^2.
\end{equation} 
For the electron-photon system the second, photon-related variable is required. The most natural choice \cite{TokatlyPrl2013,Ruggenthaler2etal2014} is the expectation value $P$ of the photonic coordinate  $p$
\begin{equation}\label{P}
P=\int \d p (|\psi_1(p;t)|^2+|\psi_2(p;t)|^2)p.
\end{equation}

The next step is to find the equations  which relate the basic observables, the density $n_i$ and  the field average  $P$, to  the ``external potentials'', the on-site potential $v_i$ and the external current $J_{ex}$. Therefore we proceed to deriving the equations of motion for the two fundamental variables.

In order to derive the relevant equation of motion for $n_i$ we calculate the time-derivative of \eqref{D1} and then substitute the derivatives of the wave function from the \SE \eqref{SE1}. The result takes the following form
\begin{equation} \label{CE1}
\dot n_1(t)=- 2 \Im[ T \rho_{1 2}(t)]
\end{equation}
where $\dot n = \partial_t n$, and $\rho_{1 2}(t)$ is the density matrix ,
\begin{equation} \label{rho1}
\rho_{1 2} =\int \rho_{1 2}(p;t) \d p=\int \psi_1^*(p;t)\psi_2(p;t)\d p.
\end{equation}
The conservation of the particles dictates that the change in the density in one site is equal to minus the change in the other site $\dot n_2= 2 \Im T\rho_{1 2}$. Obviously Eq.~(\ref{CE1}) is a lattice version of the continuity equation for site 1.
Since in the left hand side of Eq.~(\ref{CE1}) we have the time derivative of the on-site density, the right hand side should be identified with a  current flowing along the link connecting the two sites
\begin{equation} \label{CR1}
J_{1 2}(t)=2 \Im [T \rho_{1 2}(t)].
\end{equation}

Differentiating the continuity equation \eqref{CE1} with respect to time and replacing the derivative of the wave function from the \SE we get an equation which connects  the on-site density $n_i$ to the on-site potential $v_i$
\begin{eqnarray} \label{LF1}
\ddot n_1(t)&=& -2T\Big(  \Re[\rho_{1 2}](v_1(t)-v_2(t)) \nn\\
 &&+T (n_1-n_2)+2\lambda \int \Re[\rho_{1 2}(p) 	] p \d p\Big).
\end{eqnarray}
Physically this equation can be interpreted as the (discrete) divergence of the force balance equation for the two-site model \cite{FarTok2012}.
 
A special role of Eq.~\eqref{LF1} for TDDFT follows from the fact that it explicitly relates the potential disbalance   $v_1(t)-v_2(t)$ to the density $n_i(t)$ and its derivatives. 
Like before, the conservation of the particle imposes $\ddot n_2=-\ddot n_1$.  Hence the force balance equation for $n_2(t)$ is obtained from \Q~\eqref{LF1} by changing the sign in the right hand side.

It is worth noting that the coefficient of the potential disbalance, $v_1(t)-v_2(t)$, in the force balance equation \eqref{LF1} is the kinetic energy $k_{12}=2T \Re [\rho_{12}]$ therefore for the current $J_{12}$ and kinetic $k_{12}$ we have:
\begin{equation}\label{k+j}
K_{12}+\id J_{12}=2T \rho_{12}
\end{equation}  

Importantly, Eq.~\eqref{LF1} contains only the potential disbalance $v_1-v_2$ which reflects the well known gauge redundancy of TDDFT.  For a given density the force balance equation fixes the on-site potential up to a constant.  In order to resolve this issue we fix the gauge by considering on-site potentials which sum up to zero 
\begin{equation}\label{ZeroSum}
v_1=-v_2=v
\end{equation} 
This can be interpreted as a switching from the whole two dimensional space of all allowed potentials to the one dimensional space of equivalence classes for physically distinct potentials.

Next, we need to derive a similar equation for $P$. So we differentiate \eqref{P} with respect to $t$ and simplify the right hand side using the \SE \eqref{SE1} and the result is as follows:
\begin{equation}\label{P-dot}
\dot P=  \Im \big[\int \Big(  \psi_1 ^*\partial_p \psi_1 + \psi_2 ^* \partial_p \psi_2 \Big)\d p\big]
\end{equation}
For brevity we suppressed the explicit $p$- and $t$-dependence of the wave function.

By differentiating Eq.~(\ref{P-dot}) with respect to time and again substituting the time derivatives from the \SE  \eqref{SE1} we get an equation which relates $J_{ex}$ to $P$, $\ddot P$ and $n_i$ 
\begin{eqnarray}\label{P-eq}
\ddot P =- \omega^2 P - \lambda(n_1-n_2)-J_{ex}(t).
\end{eqnarray}
This equation is, in fact, the inhomogeneous Maxwell equation projected on the single photon mode \cite{Ruggenthaler2etal2014}.

In the next subsection we will use the force balance equation (\ref{LF1}) and equation of motion \eqref{P-eq} to analyze the existence of a TDDFT-like  theory for a two-site lattice coupled to a photonic field.

%##################################################################################
\subsection*{Statement of the mathematical problem and the basic existence theorem}%##############
%###################################################################

The standard TDDFT is based on the existence a one-to-one map between the time-dependent density and the external potential. In the case of the electron-photon system the map is slightly different. Here the two basic observables, the on-site density $n_i$ and the expectation value of the field $P$, are mapped to the two external fields, the on-site potential $v(t)$ and the external current $J_{ex}(t)$. 

Equations \eqref{SE1}, \eqref{D1} and \eqref{P} uniquely determine the instantaneous wave function $\Psi(t)$, the on-site density $n_i$ and the field average $P$ as functionals of the initial state $\Psi_0$ , the on-site potential $v(t)$ and the external current $J_{ex} (t)$. This defines a unique direct map $\{\Psi_0, v,J_{ext}\}\to \{\Psi, n, P\}$.

The TDDFT formalism relies on the existence of a unique inverse map: $\{\Psi_0, n, P \}\to \{\Psi,v,J_{ext}\}$.  To prove the inverse map we follow the NLSE approach to TDDFT-type theories \cite{Tokatly2007,Maitra2010,TokatlyUni2011,TokatlyL2011,FarTok2012}.

Assuming $n_i(t)$ and $P(t)$ are given  functions of time, we express  $v$ and $J_{ex}$ from  the equations \eqref{LF1} and \eqref{P-eq} as follows
\begin{eqnarray}
&&v=- \frac{\ddot n_1+2T^2 (n_1-n_2)+4T\lambda\int \Re[\rho_{1 2}(p) ] p \d p}{ 4T \Re[\rho_{1 2}]}
\label{V1}\\
&&J_{ex}= -\ddot P - \omega^2 P -\lambda(n_1-n_2)
\label{J_ex1}
\end{eqnarray}
where we assumed that $\Re[\rho_{12}]\neq0$.
 
We note that at any time, including the initial time $t=t_0$ , the given density has to be consistent with the wave function. At $t=t_0$ this means that the right hand sides of \Q \eqref{D1} evaluated at the initial wave function $\Psi_0$ has to be equal to  $n_i(t_0)$ in the left hand side. The same follows for the first derivative of the density $\dot n_i$ and the field average $P$ and its first derivative $\dot P$. All of them should to be consistent with the initial state $\Psi_0$ trough Eqs.~\eqref{CE1}, \eqref{P} and \eqref{P-dot} respectively. The consistency conditions which should be fulfilled are the following
\begin{subequations}\label{ConE1}
\begin{eqnarray}
&&n_i(t_0)= \int \d p |\psi_i(p;t_0)|^2,\\
&&\dot n_1(t_0)=- 2 \Im[ T \rho_{1 2}(t_0)]\\
&&P(t_0)=\int \d p~ p\big(|\psi_1(p;t_0)|^2+|\psi_2(p;t_0)|^2\big)\\
&&\dot P(t_0)= \Im \Big[\int \d p \Big(  \psi_1 ^*(p;t_0)\partial_p \psi_1(p;t_0) \nn\\
&&~~~~~~~~~~~~+ \psi_2 ^*(p;t_0) \partial_p \psi_2(p;t_0) \Big)\Big]
\end{eqnarray}
\end{subequations}

The on-site potential $v$ of Eq.~\eqref{V1} and  the external current $J_{ex}$ of Eq.~\eqref{J_ex1} can be substituted as a functionals of $n$, $P$ and $\Psi$  in to the  the \SE \eqref{SE1}. The  result is a universal NLSE in which the Hamiltonian is a function of the instantaneous  wave function and the (given) basic variables
\begin{equation}\label{NLSE1}
\id \partial_t \Psi(t)= H[n,P,\Psi]\Psi(t).
\end{equation}
Now the question of existence of a unique QED-TDDFT map $\{\Psi_0, n_i, P \}\to \{\Psi,v,J_{ext}\}$ can be formulated mathematically as the problem of existence of a unique solution to NLSE \eqref{NLSE1} with given $n_i (t)$, $P(t)$ and $\Psi_0$.   

\textit{Theorem 1. (existence of QED-TDDFT for a Hubbard dimer coupled to a photonic mode) } Assume that the on-site density $n_i(t)$ is a positive, continuous function of time, which has a continuous second derivative and add up to unity, $n_1(t)+n_2(t)=1$. Consider $P(t)$ which is a continuous function of time with a continuous second derivative. 
Let $\Omega$  be a subset of the Hilbert space where $\Re[\rho_{12}]\neq 0$. If the initial state $\Psi_0\in \Omega$ , and the consistency conditions of Eqs.~\eqref{ConE1} hold true, then:

%\noindent
(\textit{i}) there is an interval around $t_0$ in which NLSE \eqref{NLSE1} has a unique solution and, therefore, there exists a unique map $\{\Psi_0, n_i, P \}\to \{\Psi,v,J_{ext}\}$.

%\noindent
(\textit{ii}) The solutions (i.~e. the QED-TDDFT map) is not global in time if at some $t^*>t_0$ the boundary of $\Omega$ is reached.

\textit{Proof}: By the condition of the theorem $\Psi_0 \in \Omega $ where $\Re[\rho]\neq 0$. Therefore  the on-site potential $v$ can be expressed in terms of the density and the wave function as given by \eqref{V1}, and the Hamiltonian $\hat H [n,P,\Psi]$ in the universal NLSE is well defined.

Let us rewrite NLSE \eqref{NLSE1} in the following form
\begin{equation}\label{NLSE2}
\id \partial_t \Psi=\left( \hat H_0+\hat H_1[n,P,\Psi] \right)\Psi
\end{equation} 
where $\hat H_0$ is the time-independent (linear) part of the Hamiltonian,
\begin{equation} \label{H_0}
\hat H_0=- \frac{1}{2} \partial _p ^2+\frac{1}{2} \omega ^2 p^2 +\lambda p \hat \sigma_z - T  \hat \sigma_x,  
\end{equation}
and $\hat H_1$ contains all time-dependent, in particular non-linear, terms, 
\begin{equation}\label{H_1}
\hat H_1[n,P,\Psi]=J_{ex}[n,P] p +v[n,P,\Psi] \hat \sigma_z .
\end{equation}
Here $J_{ex}[n,P]$ and $v[n,P,\Psi]$ are defined by Eqs.~(\ref{J_ex1}) and (\ref{V1}), respectively.

Since $\hat H_0$ is the Hamiltonian of the static shifted harmonic oscillator it defines a continuous propagator in the Hilbert space of square integrable functions. Therefore Eq.~\eqref{NLSE2} can be transformed to the following integral equation,
\begin{eqnarray}\label{nl-Int}
\Psi(t)=&&e^{-\id \hat H_0 (t-t_0)} \Psi_0 \\
&&-\id \int_{t_0} ^t e^{-\id \hat H_0 (t-s)}\hat H_1[n(s),P(s),\Psi(s)] \Psi(s) \d s\nn.
\end{eqnarray}
To prove the existence of solutions to this equation we can use well established theorems from the theory of quasilinear PDE  \cite{Segal1963,Haraux}. In particular, we apply the the following result. Consider an integral equation of the form, 
\begin{eqnarray}\label{nl-Int2}
u(t)=  W(t,t_0) u_0+ \int_{t_0} ^t W(t,s) K_s(u(s))\d s,
\end{eqnarray}
where $W(t,s)$ is a continuous linear propagator on $T= [t_0,\infty)$ and the kernel $K_t(u)$ is continuous function of time, which is locally Lipschitz in a Banach space $\mathcal{B}$. Then there is an interval $[t_0,t^*)$ where \Q~\eqref{nl-Int2} has a unique continuous solution. 

In our case we consider $L^2$ as the proper Banach space $\mathcal{B}$.  The kernel $K_t(\Psi)=\hat H_1[n(t),P(t),\Psi] \Psi$ in Eq.~(\ref{nl-Int}) is continuous and Lipschitz in $L^2$ if $n(t)$, $\ddot n(t)$, $P(t)$ and $\ddot P(t)$ are continuous functions of time,  $\Psi \in \Omega$, and the consistency conditions Eqs.~\eqref{ConE1} are fulfilled. Hence if all conditions of the theorem are satisfied Eq.~\eqref{nl-Int} has a unique solution. Moreover since in this case $\Psi_0$ is in the domain of $H_0$, $\Psi_0\in D(H_0)$, there exists a unique differentiable (strong) solution of Eq.~ \eqref{nl-Int} which proves the statement $(i)$ of the theorem.

The extension theorems for quasilinear PDE imply that the local solution can not be extended beyond some maximal existence time $t^* > t_0$ only in two cases: first, at  $t \to t^*$ the solution becomes unbounded or, second, at $t \to t^*$ it reaches the boundary of $\Omega$. In our case the solution is guaranteed to be normalized and thus bounded. Therefore we are left only with the second possibility, which proves the statement $(ii)$ and completes the proof of the theorem.

The above theorem generalizes the results of Ref.~\onlinecite{Ruggenthaler2etal2014} where the uniqueness (but not the existence) of the map $\{\Psi_0, n_i, P \}\to \{\Psi,v,J_{ext}\}$ has been proven for analytic in time potentials using the standard Taylor expansion technique.

The Theorem~1 can be straightforwardly generalized to the case of multiple photon modes. The only difference is that $\hat H_0$ in Eq.~(\ref{H_0}) becomes the Hamiltonian of a multidimensional shifted harmonic oscillator. The rest of the proof remains unchanged. This proves the existence of QED-TDDFT for the spin-boson model in its standard form \cite{Leggettetal1987}. A less obvious generalization for the system of many interacting electrons on a many-site lattice in is presented in the next section.  

\section{QED-TDDFT for many electrons on many-site lattices interacting with cavity photons} \label{many-electron-photon}

In the previous section we proved the QED-TDDFT existence theorem for a system of one electron on a two-site lattice coupled to a photonic mode. Below we generalize our results to the case of $N$ interacting electrons on a $M$-site lattice  coupled to  $L$ photonic modes. The state of the system is described by an electron-photon wave function $\psi(\rv_1, \cdots, \rv_N;\{p\})$ where coordinates $\rv_i$ of the particles $(i=1,2,\cdots,N)$ take values on the lattice sites and $\{p\}$ is the set of continuous coordinates describing the photonic (oscillatoric) degreed of freedom $\{p\}=\{p_1,p_2,...,p_L\}$. Again we assume that the electronic   subsystem is driven by  classical on-site potentials $v(\rv;t)$ and each photonic mode is coupled to corresponding external current $J_{ex}^\alpha(t)$. As usual, assuming that the size of the lattice is much smaller than the wave length of the photon field, we describe the electron-photon coupling at the level of the dipole approximation with $\boldsymbol{\lambda}_\alpha$ being the coupling constant to the $\alpha$-photon. 

The following time-dependent  \SE describes the time evolution of the system from the initial state $\psi_0(\rv_1 \cdots \rv_N;\{p\})$ 
\begin{eqnarray}\label{SE-mb1}
&&\id \partial_t \psi(\rv_1,...,\rv_N;\{p\})=-\sum_{i=1}^{N}\sum_{\xv_i}T_{\rv_i,\xv_i}\psi( ...,\xv_i,...;\{p\})\nn\\
&+&\sum_{i=1}^Nv(\rv_j ;t)\psi(\rv_1,...,\rv_N;\{p\})+\sum_{j > i} w_{\rv_i , \rv_j} \psi(\rv_1,...,\rv_N;\{p\})\nonumber\\
&+&\sum_{\alpha=1}^K \Big[ -\frac{1}{2} \partial^2 _{p}+\frac{1}{2} \omega^2 _\alpha p_\alpha ^2+J_{ex}^\alpha(t)p_\alpha\Big]\psi(\rv_1,...,\rv_N;\{p\})\nn\\
&+& \sum_{i=1}^N\sum_{\alpha=1}^K \boldsymbol{\lambda}_\alpha \cdot \rv_i  p_\alpha\psi (\rv_1,...,\rv_N;\{p\})
\end{eqnarray}
where the real coefficients  $T_{\rv,\rv'}=T_{\rv',\rv}$ correspond to the rate of hopping from site $\rv$ to site $\rv'$ (for definiteness we set $T_{\rv,\rv}=0$), and $w_{\rv,\rv'}$ is the potential of a pairwise electron-electron interaction.

Following the logic of Sec.~\ref{two-level} we define the on-site density $n(\rv)$ and the expectation value of the field $P_\alpha$ for a mode $\alpha$, which are  the basic variables for the QED-TDDFT
\begin{eqnarray}
n(\rv)&=& N  \sum_{\rv_2, ... ,\rv_N} \int |\psi(\rv,\rv_2,...,\rv_N;\{p\})|^2 \d \pv ,\label{D-mb1}\\
P_\alpha &=&  \sum_{\rv_1,...,\rv_N} \int p_\alpha|\psi(\rv,...,\rv_N;\{p\})|^2 \d \pv\label{P-mb1}
\end{eqnarray} 
where $\d \pv=\d p_1\cdots \d p_L$. 

Similarly to the two-site case we derive the force balance equation by calculating the second derivative of the density \eqref{D-mb1} and using the \SE \eqref{SE-mb1} to simplify the terms with the time derivative of the wave function 
\begin{equation}\label{FB-mb1}
\ddot n(\rv)=2 \Re \sum_{\rv'} T_{\rv,\rv'} \rho(\rv,\rv')(v(\rv';t)-v(\rv;t) ) +q(\rv;t)+f(\rv;t)
\end{equation}
where $q(\rv;t)$ is the lattice divergence of the internal forces
\begin{eqnarray}
q(\rv;t)=&-&2\Re\sum_{\rv',\rv''} T_{\rv,\rv'}\Big[ T_{\rv',\rv''}\rho(\rv,\rv'' )-T_{\rv,\rv''}\rho(\rv',\rv'' )\nn\\
&+&\rho_2(\rv,\rv'',\rv')(w_{\rv,\rv''}-w_{\rv',\rv''})\Big],
\end{eqnarray}
and $f(\rv;t)$ is  the force exerted on electrons from the photonic subsystem 
\begin{equation}
f(\rv;t)=2\Re\sum_\alpha\sum_{\rv'} T_{\rv,\rv'}\boldsymbol{\lambda}_\alpha \cdot(\rv'-\rv) \int p_\alpha \rho(\rv,\rv';p_{\alpha})\d p_\alpha .
\end{equation} 
Here $\rho(\rv,\rv')$ is the one-particle density matrix 
\begin{eqnarray}
&&\rho(\rv,\rv')=\int  \rho_{}(\rv,\rv';p_{\alpha})\d p_\alpha\\
&&=N\sum_{\rv_2, ... ,\rv_N} \int \psi^*(\rv,\rv_2...,\rv_N;\{p\}) \psi(\rv',\rv_2,...,\rv_N; \{p\})\d  \pv.\nn
\end{eqnarray}
and $\rho_2(\rv,\rv'',\rv')$ is the two-particle density matrix
\begin{eqnarray}
\rho_2(\rv,\rv'',\rv')=&&N(N-1)\sum_{\rv_3, ... ,\rv_N} \int \big[ \psi^*(\rv,\rv''...,\rv_N;\{p\})\nn\\
&& \times\psi(\rv',\rv'',...,\rv_N; \{p\})\big] \d \pv
\end{eqnarray}

The equation of motion for the field average $P_\alpha$ \eqref{P-mb1}  is derived in the same manner as in Sec.~\ref{two-level} by calculating the second time derivative of $P_\alpha$ and using the \SE \eqref{SE-mb1},
\begin{equation}\label{P-mb-eq1}
\ddot P_\alpha=- \omega^2 _\alpha P_\alpha-J_{ex}^\alpha-\boldsymbol{\lambda}_\alpha \cdot \dv
\end{equation}
where $\dv$ is the total dipole moment of the $N$-electron system
\begin{equation}
\dv=\sum_{\rv} \rv n(\rv).
\end{equation}

The existence of QED-TDDFT is equivalent to the existence of the inverse map  $\{\Psi_0, n(\rv), P_\alpha \}\to \{\Psi,v,J_{ext}^\alpha\}$. To study this map we compile the universal NLSE by expressing the on-site potential $v(\rv)$ and the external current $J_{ex}^\alpha$ in terms of the fundamental observables  $n(\rv)$ and $ P_\alpha$, and the wave function $\Psi$. 

To find the current $J_{ext}^\alpha$ as a functional of the field average $P_\alpha$ and the density $n(\rv)$ we only need to rearrange \Q~\eqref{P-mb-eq1}
\begin{equation}\label{P-mb-eq2}
J_{ex}^\alpha=-\left(\ddot P_\alpha+ \omega^2 _\alpha P_\alpha+\boldsymbol{\lambda}_\alpha \cdot \dv\right).
\end{equation}
The problem of finding the potential $v(\rv)$ as a functional of $n(\rv)$ and $\Psi$ is more involved \cite{FarTok2012} as we need to solve the system of $M$ linear equations, \Q~\eqref{FB-mb1}, for $v(\rv)$. Let us first rewrite \eqref{FB-mb1}  in a matrix form as follows
\begin{equation}\label{FB-mb2}
\hat{K}[\Psi]V= S[\ddot n,\Psi],
\end{equation}
where $\hat{K}$ is a real symmetric $M\times M$ matrix with elements 
\begin{equation} \label{K1}
k_{\rv,\rv'}[\Psi]=2\Re\left[T_{\rv,\rv'}\rho(\rv,\rv')-\delta_{\rv,\rv'}\sum_{\rv''}T_{\rv,\rv''}\rho(\rv,\rv'')\right].
\end{equation}
and $V$ and $S$ are $M$-dimensional vectors with components
\begin{subequations}\label{VandS}
\begin{eqnarray}
&v_{\rv}=v(\rv),&\\
 &s_{\rv}[\ddot n,\Psi]=- \ddot n(\rv)-q[\Psi](\rv)-f[\Psi](\rv)&.
\end{eqnarray}
\end{subequations} 

The problem of inverting (solving for $v(\rv)$) the force balance equation \eqref{FB-mb2} for a general lattice has been analyzed in Ref. [\onlinecite{FarTok2012}] in the context of the standard electronic TDDFT.  The same argumentation regarding the properties of the matrix $\hat K$ is applicable in the present case. Solving Eq.~\eqref{FB-mb1} for the on-site potential $v(\rv)$ is equivalent to multiplying both sides of \Q \eqref{FB-mb2} by inverse of the $\hat K$-matrix. Therefore the matrix $\hat K$ must be non-degenerate. At this point it is worth noting that because of the gauge invariance $\hat{K}$ matrix  \eqref{K1} always has at least one zero eigenvalue that corresponds to a space-constant eigenvector. Therefore if $\mathcal{V}$ is the $M$-dimensional space of lattice potentials $v(\rv)$, then the invertibility/nondegeneracy of $\hat{K}$ should always refer to the invertibility in an $M-1$-dimensional subspace of $\mathcal{V}$, which is orthogonal to a constant vector $v_C(\rv)=C$. In more physical 
terms this means that the force balance equation \eqref{K1} determines the self-consistent potential $v[n,\Psi](\rv)$ only up to an arbitrary constant, like \Q \eqref{LF1} in section \ref{two-level}. Therefore in the orthogonal subspace subspace of $\mathcal{V}$ we have 
\begin{equation}\label{V-mb1}
V=\hat K ^{-1} S
\end{equation}
where by $\hat K ^{-1}$ we mean inversion in the subspace of $\mathcal{V}$. Equation~(\ref{V-mb1}) the required functional $v[n,\Psi](\rv)$ which can be used to construct the universal NLSE.

Finally, like in the two-site case (see Sec.~\ref{two-level}) the initial sate $\Psi_0$, the density $n(\rv)$  and the field average $P_\alpha$ should satisfy the consistency conditions at $t=t_0$ 
\begin{subequations}\label{Con-mb1}
\begin{eqnarray}
&&n(\rv;t_0)=  N  \sum_{\rv_2, ... ,\rv_N} \int |\psi_0(\rv,..., \rv_N;\{p\})|^2 \d \pv,\\
&&\dot n(\rv;t_0)=- 2 \Im[\sum_{\rv'} T_{\rv,\rv'} \rho_0(\rv,\rv')]\\
&&P_\alpha(t_0)= \sum_{\rv_1,...,\rv_N} \int p_\alpha|\psi_0(\rv_1,...,\rv_N;\{p\})|^2 \d \pv\\
&&\dot P_\alpha(t_0)=  \Im\sum_{\rv_1,...,\rv_N}  \int \big[ \psi_0^*(\rv_1,..., \rv_N;\{p\})\nn\\
&&~~~~~~~~~ \times \partial_{p_\alpha} \psi_0(\rv,..., \rv_N;\{p\})\big]\d \pv
\end{eqnarray}
\end{subequations}

By substituting the on-site potential $v(\rv)$ from \Q \eqref{V-mb1}, and the field average $P_\alpha$ from \Q \eqref{P-mb-eq2} in to the \SE \eqref{SE-mb1} we find the proper NLSE
\begin{equation}\label{NLSE-mb1}
\id \partial_t \Psi(t)= H[n,P,\Psi]\Psi(t)
\end{equation}
which we use to prove the existence of the unique inverse map $\{\Psi_0, n(\rv), P_\alpha \}\to \{\Psi,v(\rv),J_{ext}^\alpha\}$ and thus the existence of the QED-TDDFT in a close analogy with the Theorem~1. 

{\it Theorem 2. (existence of the QED-TDDFT for lattice systems coupled to cavity photons)} --- Assume that a given time-dependent density $n(\rv;t)$ is nonnegative on each lattice site, sums up to the number of particles $N$, and has a continuous second time derivative $\ddot n(\rv;t)$. Also assume that $P_\alpha(t)$ is a continuous function of $t$ and has a continuous second time derivatives $\ddot P_\alpha(t)$. Let $\Omega$ be a subset of the $N$-particle Hilbert space $\mathcal{H}$ where the matrix $\hat{K}[\Psi]$ \eqref{K1} has only one zero eigenvalue corresponding to the space-constant vector. If the initial state $\Psi_0\in\Omega$, and at time $t_0$ the consistency conditions of Eq.~\eqref{Con-mb1} are fulfilled, then:

\noindent
$(i)$ There is a time interval around $t_0$ where the many-body NLSE  \eqref{NLSE-mb1} has a unique solution that defines the wave function $\Psi(t)$ and the external potentials, $v(t)$and $J_{ex} ^\alpha(t)$, as unique functionals of the density $n(\rv;t)$, field average $P_\alpha$, and the initial state $\Psi_0$;

\noindent
$(ii)$ The solution of item $(i)$ is not global in time if and only if at some maximal existence time $t^*>t_0$ the boundary of $\Omega$ is reached. 

The proof of this theorem goes along the same lines as the proof of Theorem~1 in Sec.~\ref{two-level} .  We transform NLSE \eqref{NLSE-mb1} to a multidimensional integral equation similar to \eqref{nl-Int} and then apply the general existence results \cite{Segal1963,Haraux} for equations of the type of Eq.~(\ref{nl-Int2}) to show that the  statements (\textit{i}) and (\textit{ii}) are in fact true. We skip the details as the procedure is mostly a straightforward repetition of the proof presented in Sec.~\ref{two-level}.

%%%%%%%%%%%%%%%%%%%%%%%%%%%%%%%%%%%%%%%%%%%%%%%%%%%%%%%%%%%%%%%%%%%%%%%%%%%%%%%%%%%%%
\section{Time-dependent $v$-representability for a system evolving from the ground state}\label{ground-state-theorem}
%%%%%%%%%%%%%%%%%%%%%%%%%%%%%%%%%%%%%%%%%%%%%%%%%%%%%%%%%%%%%%%%%%%%%%%%%%%%%%%%%%%%%

In this section we will show that the ground state of a quite general lattice Hamiltonian  belongs to the $v$-representability subset $\Omega$. This implies  the map  $\{\Psi_0, n, P_\alpha \}\to \{\Psi,v,J_{ext}^\alpha\}$ is guaranteed to exist if the dynamics starts from the ground state. The main theorem of this section is a generalization of Theorem~2 in Ref.~\onlinecite{FarTok2012}.

Consider the following the lattice Hamiltonian of  many mutually interacting electrons coupled to photonic modes 
\begin{equation}\label{H1}
\hat H_{\alpha,\beta,\gamma}= (\hat T + \hat V+\alpha  \hat W_{e-e})\otimes \unit_{ph}+ \unit_e \otimes \hat H_{ph}+\beta	 H_{e-ph}
\end{equation}
where $\hat T$ is the usual lattice operator of the kinetic energy, $\hat V$ corresponds to the interaction with a local external potential, $\hat W$ describes the electron-electron interaction, $\hat H_{ph}$ is the photonic Hamiltonian and $H_{e-ph}$ is the the Hamiltonian for the interaction between electrons and the photon modes, $\unit_{ph}$ and $\unit_e$ are, respectively, the unit matrices in the photonic and the electronic sectors of the Hilbert space, and $\alpha$ and $\beta$ are  real coefficients.

Here we demonstrate that the ground state of the Hamiltonian \eqref{H1}, for any $\alpha,\beta \in \mathbb{R} $ and any on-site potential, belongs to the $v$-representability subset $\Omega$ if all terms in Eq.~(\ref{H1}) except $\hat T$ commute with the density operator $\hat n_\rv$.  The proof of this quite general statement closely follows the proof of the Theorem~2 in Ref. [\onlinecite{FarTok2012}].  Therefore below we only briefly go through the main line of arguments.  

Assume that $\Psi_k=|k\rangle$ form a complete set of eigenstates for the Hamiltonian \eqref{H1} and let $\Psi_0=|0\rangle$ be the ground state. We will show that the matrix $\hat{K}[\Psi_0]$ evaluated at the ground state is strictly negative definite in the subspace of potentials that are orthogonal to a space-constant vector $V_{C}$. That is,
\begin{equation}
 \label{K-negativity1}
 V^{T}\hat{K}[\Psi_0]V \equiv \sum_{\rv,\rv'}v({\rv})k_{\rv,\rv'}v({\rv}) < 0,
\end{equation}
for all $M$-dimensional vectors $V=\{v({\rv})\}$ which are orthogonal to the spatially constant potential
\begin{equation}
 \label{v-ortho}
 V^{T}V_{C} = C\sum_{\rv}v({\rv})=0,
\end{equation}
where $V^T$ stands for a transposed vector. Therefore $\hat K[\Psi_0]$ is nondegenerate in the subspace orthogonal the to constant potentials.

Using the $f$-sum rule and the spectral representation of the density-density response function (see, for example, Ref.~\onlinecite{VignaleBook}) one can represent the elements of $\hat K$-matrix Eq.~\eqref{K1} as follows (see Ref.[\onlinecite{FarTok2012}] for details)
\begin{equation}\label{Kground}
k_{\rv,\rv'}[\Psi_0]= -4\Re\sum_{k} \omega_{k0}
\langle 0| \hat{n}_{\rv}| k \rangle \langle k| \hat{n}_{\rv'}| 0\rangle.
\end{equation}
where  $\omega_{k0}=E_k-E_0$ is excitation energy of the system from the ground state to the state $k$. 

Substituting  $k_{\rv,\rv'}$ of  Eq.~(\ref{Kground}) into the left hand side of Eq.~(\ref{K-negativity1}) we find  the following result
\begin{eqnarray}
\nonumber
V^{T}\hat{K}[\Psi_0]V &=& - 4\sum_{k}\omega_{k0}{\Big |}\sum_{\rv} v({\rv})\langle 0|\hat{n}_{\rv}|k\rangle{\Big |}^2\\
\label{K-negativity2}
 &=&-4\sum_{k}\omega_{k0}|\langle 0|\hat{v}|k\rangle|^2 \le 0,
\end{eqnarray}
where $\hat{v}$ is  an operator corresponding to the potential $v({\rv})$,
\begin{equation}
 \label{v-operator1}
 \hat{v} = \sum_{\rv} v({\rv})\hat{n}_{\rv}.
\end{equation}

The equality in Eq.~(\ref{K-negativity2}) holds only for a space-constant potential $v_C(\rv)=C$. Indeed, since each term in the sum in Eq.~(\ref{K-negativity2}) is non-negative, the result of summation is zero if and only if
\begin{equation}
 \label{v_0k}
 \langle 0|\hat{v}|k\rangle=0, \quad \text{for all $k\ne 0$}.
\end{equation}
Assuming that Eq.~(\ref{v_0k}) is fulfilled and expanding the vector $\hat{v}|0\rangle$ in the complete set of states $\{|k\rangle\}$ we get
\begin{equation}
 \label{VPsi0}
 \hat{v}|0\rangle = \sum_k |k\rangle\langle k|\hat{v}|0\rangle = |0\rangle\langle 0|\hat{v}|0\rangle \equiv \lambda|0\rangle.
\end{equation}
Therefore the condition of Eq.~(\ref{v_0k}) implies that the ground state $|0\rangle$ is an eigenfunction of the operator $\hat{v}$. Since $\hat{v}$ corresponds to a local multiplicative one-particle potential this can happen only if the potential is spatially  constant. Hence for all potentials which are orthogonal to a constant in a sense of Eq.~(\ref{v-ortho}) the strict inequality in Eq.~(\ref{K-negativity2}) takes place. This means that matrix  $\hat{K}[\Psi_0]$ is negative definite and thus invertible in the $M-1$-dimensional subspace of $\mathcal{V}$ orthogonal to a constant vector $V_C$. In other words, the ground state of $N$-particle system on a connected lattice does belong to the $v$-representability subset $\Omega$. This result combined with the general existence theorem of Sec.~\ref{many-electron-photon} proves the following particular version of the time-dependent $v$-representability theorem.

{\it Theorem 3.} --- Let the initial state $\Psi_0$ for a time-dependent many-body problem on a connected lattice corresponds to a ground state of a Hamiltonian of the form \eqref{H1}. Consider  continuous positive density $n(\rv;t)$ and field average $P(t)$ which satisfy the consistency conditions of Eqs.~\eqref{Con-mb1} and has a continuous second time derivative. Then  there is a finite interval around $t_0$ in which  $n(\rv;t)$ and $P(t)$ can be  reproduced uniquely by a time evolution of \SE \eqref{SE-mb1} with  some time dependent on-site potential $v_i(t)$ and  external current $J_{ex}(t)$. 

Note that Theorem~3 is valid for any Hamiltonian of the form of Eq.~\eqref{H1} as long as all the terms in the  Hamiltonian,  except the kinetic part, commute with the density operator  $\hat n_\rv$, and, therefore, do not contribute to the $\hat K$-matrix.  An important special case is when the initial state is the interacting many-electron ground state which is decoupled from the photonic field, $\beta=0$. In this case  the ground state is a direct product of the electronic ground state and photonic ground state. Another practically relevant case of $\alpha=\beta=0$ corresponds to the initial state in a form of the direct product of the noninteracting many-electron wave function (the Slater determinant) and the photonic vacuum. For all those cases the local $v$-representability is guarantied by the above Theorem~3.

%########################################
\section{Conclusion}\label{conclusion}
%########################################

In this work we extended the formalism of the lattice TDDFT \cite{FarTok2012} and presented a rigorous proof of the mapping theorem of QED-TDDFT for many-electron lattice systems coupled to quantized photonic modes. First we considered the simplest non-trivial model of a one-electron Hubbard dimer coupled to a single photonic mode, which is identical to the Rabi and spin-boson models. We identified a pair of basic variables describing the electronic and photonic degrees of freedom, and showed that the existence of the QED-TDDFT map is equivalent to the existence of a unique solution to a certain system of non-linear partial differential equations (the universal NLSE). We proved that the Cauchy problem for this NLSE indeed has a unique solution and, therefore, the unique QED-TDDFT map exists, provided the basic variables have a continuous second time derivative and the initial state belongs to some well defined subset of the Hilbert space (the $v$-representability subset). Further we generalized the theory to many 
electrons and multiple photonic modes. We proved that the same QED-TDDFT mapping can be constructed for this generic interacting electron-photon system. Finally, we showed that the ground state of a quite general electron-photon lattice Hamiltonian belongs to the $v$-representability subset $\Omega$ of the Hilbert space. Therefore if the system evolves from such ground state the local $v$-representability is always guaranteed. 

The main difference and, in fact, the main mathematical difficulty of the present theory, as compared to the purely electronic lattice TDDFT \cite{TokatlyL2011,FarTok2012}, is the existence continuum variables describing photonic modes. As a result the Hilbert space becomes an infinite dimensional functional space, and the universal NLSE turns into a system of PDE. In a certain sense the lattice QED-TDDFT is on half way between the electronic lattice TDDFT and the usual TDDFT in the continuum space. Hopefully the present rigorous results will shed new light on remaining unresolved issues of TDDFT and thus deepen our understanding of this popular and practically important theory.

\section*{Acknowledgements}
We acknowledge financial support from the Spanish Ministry of Economy and Competitiveness (Grant No. FIS2013-46159-C3-1-P) and Grupos Consolidados UPV/EHU del Gobierno Vasco (Grant No. IT578-13).

%SpinBosonModel
%\bibliographystyle{unsrt}
%\bibliography{tddft,PDE-ref,SpinBosonModel,Cavity-QED,Math-ref}
%Merlin.mbs v4.21 2009-07-09.
%
\end{document}